\documentclass[sigconf]{acmart}

\usepackage{balance}

\usepackage{amsmath,amsthm,amssymb, color}
\usepackage{booktabs} 
\usepackage{comment}
\usepackage{subcaption}
\usepackage{array}
\usepackage[]{algorithm}
\usepackage[noend]{algpseudocode}
\usepackage{tabularx}
\usepackage{hyperref}       
\usepackage{url}            
\usepackage{amsfonts}       
\usepackage{nicefrac}       
\usepackage{microtype}      
\usepackage{mathtools}
\usepackage{multirow}
\usepackage{setspace}
\usepackage{graphicx}
\usepackage{caption}
\usepackage{subcaption}
\usepackage[titletoc,title]{appendix}
\usepackage{afterpage}
\usepackage{float}
\usepackage{xcolor, colortbl} 
\definecolor{LightCyan}{rgb}{0.88 , 1, 1}
\definecolor{Gray}{gray}{0.85}

\usepackage{pgfplots}
    \usepackage{tikz}
\usetikzlibrary{positioning,chains,fit,shapes,calc}
\usepackage{tkz-graph}
\usetikzlibrary{decorations.text}
\usetikzlibrary{decorations.pathmorphing}

\newcolumntype{L}{>{\arraybackslash}m{2.8cm}}
\newcolumntype{R}{>{\arraybackslash}m{2.8cm}}

\DeclareMathOperator*{\argmin}{argmin}
\DeclareMathOperator*{\argmax}{argmax}
\DeclarePairedDelimiterX{\norm}[1]{\lVert}{\rVert}{#1}
\DeclarePairedDelimiter{\ceil}{\lceil}{\rceil}

\theoremstyle{remark}

\numberwithin{equation}{section}

\graphicspath {{../figures/}}

\def\BibTeX{{\rm B\kern-.05em{\sc i\kern-.025em b}\kern-.08emT\kern-.1667em\lower.7ex\hbox{E}\kern-.125emX}}
    
\copyrightyear{2019}
\acmYear{2019}
\setcopyright{rightsretained}
\acmConference[KDD '19]{The 25th ACM SIGKDD Conference on Knowledge Discovery and Data Mining}{August 4--8, 2019}{Anchorage, AK, USA}
\acmBooktitle{The 25th ACM SIGKDD Conference on Knowledge Discovery and Data Mining (KDD '19), August 4--8, 2019, Anchorage, AK, USA}
\acmDOI{10.1145/3292500.3330733}
\acmISBN{978-1-4503-6201-6/19/08}

\begin{document}

\title{Seeker: Real-Time Interactive Search}

\author{Ari Biswas}
\affiliation{
  \institution{Amazon.com}
  \city{Seattle}
  \state{WA}
}
\email{aritrb@amazon.com}

\author{Thai T. Pham }
\affiliation{
  \institution{Amazon.com}
  \city{Seattle}
  \state{WA}
}
\email{phamtha@amazon.com}

\author{Michael Vogelsong}
\affiliation{
  \institution{Amazon.com}
  \city{Seattle}
  \state{WA}
}
\email{vogelson@amazon.com}

\author{Benjamin Snyder}
\authornote{Work done while at Amazon}
\affiliation{
  \institution{}
  \city{}
  \state{}
}
\email{ben.snyder@gmail.com}

\author{Houssam Nassif}
\affiliation{
  \institution{Amazon.com}
  \city{Seattle}
  \state{WA}
}
\email{houssamn@amazon.com}

\renewcommand{\shortauthors}{A. Biswas et al.}

\begin{abstract}
This paper introduces Seeker, a system that allows users to adaptively refine search rankings in real time, through a series of feedbacks in the form of likes and dislikes. When searching online, users may not know how to accurately describe their product of choice in words. An alternative approach is to search an embedding space, allowing the user to query using a representation of the item (like a tune for a song, or a picture for an object). However, this approach requires the user to possess an example representation of their desired item. Additionally, most current search systems do not allow the user to dynamically adapt the results with further feedback. On the other hand, users often have a mental picture of the desired item and are able to answer ordinal questions of the form: \lq\lq Is this item similar to what you have in mind?\rq\rq\ With this assumption, our algorithm allows for users to provide sequential feedback on search results to adapt the search feed. We show that our proposed approach works well both qualitatively and quantitatively. Unlike most previous representation-based search systems, we can quantify the quality of our algorithm by evaluating humans-in-the-loop experiments. 
\end{abstract}


%

\maketitle


\section{Introduction}

Search engines and online shopping websites maintain
indices with millions of items. Often, it is difficult for a user to
accurately describe in words what they are looking
for~\cite{Teo2016airstream}. Even if the user is able to describe
their target item effectively, large index and catalog sizes mean it
is difficult to sift through similar items efficiently.

Consider the situation in which a user is searching for a
new movie to watch. They have a mental representation of the characteristics
of the movie they would enjoy but are not acquainted with the genre
keywords, latest movies, actors or directors. Being unfamiliar with
current movie jargon, they are unable to accurately describe their
preferred movie with a traditional keyword interface, nor do they have
an example photograph. However, if we show the same user another movie
they have seen and ask them \textit{\lq\lq Is this movie similar to
  the one they have in mind?  Yes or no?\rq\rq}, people can answer
such ordinal questions with less noise than absolute judgments --
i.e. finding the exact words to describe their
choice~\cite{stewart2005absolute}.

The above scenario is not restricted to movies only. In
the case of browsing for a song on a media platform, searching for a
news article on a news website, or a dress on an online platform, the
user may not be able to accurately describe
the desired item in a traditional keyword interface. 
But users could provide
relative judgments based on what they have experienced before. For
example, answers to queries like \textit{\lq\lq Songs similar to Heroes by
  David Bowie: Yes or no?\rq\rq} or \textit{\lq\lq News similar to
  that of the Queen's involvement with Brexit: Yes or no?\rq\rq} are
easier to provide. 

In addition, traditional search
engines~\cite{SEO,AmazonRecommenderSystem} and the newer representation search
systems (described in Section~\ref{sec:literature}) are temporally
static. The engines use text or imagery as the query and respond with
a ranked list of results. This ranking is based on an estimate of
relevance to the user in their current context -- location, historical
searches etc. They do not provide the user the opportunity to adapt
and fine-tune the resulting page with additional feedback. In
traditional engines, for a given user in a given session, each query
is independent of each other. Figure~\ref{fig:setting} illustrates the
difference between traditional engines and our setting.

In this paper, we describe our system, \textbf{Seeker}, that
dynamically refines search results based on real-time interactions
with the user (in the form of likes and dislikes) within a single
search session. From a customer perspective, this system adds the
feeling of an "in-store" shopping discovery experience, with a
personal curator.  In our setting, the user scrolls through a page of
items and may "like" or "dislike" any item at any time. The data
gathered from these preferences is used to update the list of results
shown in real-time, thereby iteratively closing in on what they are
looking for. To our knowledge, Seeker is the first
interactive and dynamic search experience which enables the user to
seamlessly \emph{zoom in}, \emph{zoom out}, and \emph{pivot} by
scrolling up and down and selecting items to like and dislike in an
adaptive manner.


In this work we make the following contributions:
\begin{itemize}
\item Introduce Seeker, an interactive recommendation algorithm
  deployed at scale, which adapts to customer
  inputs in real time.
\item Propose a novel evaluation metric with humans in the loop that
  allow us to quantify the quality of our proposed algorithm and
  evaluate it against other methods. Most embedding-based representational search engines in the
  past have evaluated their systems only qualitatively rather than
  quantitatively. In our experiments, we simulate the tasks of
  searching for a particular item, and quantifiably measure progress.
\end{itemize}

The paper is organized as follows. In Section~\ref{sec:literature}, we
review related papers and search engines. In
Section~\ref{sec:setting}, we describe how we model human preferences
expressed in likes and dislikes and translate those preferences into
probability distributions over our catalog. In Section~\ref{sec:rank},
we present our adaptive algorithm for making real time
recommendations. In Section~\ref{sec:result}, we 
evaluate Seeker's results. In
Section~\ref{sec:future}, we discuss directions for
future research. In Section~\ref{sec:conclusion}, we summarize our
work.

\section{Related Work}
\label{sec:literature}

Over the last few years, there has been a growing trend of exploring
new interfaces beyond traditional keyword search, and in particular,
visual-based search~\cite{datta2008image}. In~\cite{street2shop},
users query relevant items by uploading real
world photographs of clothing. The engine then displays results that
are visually similar to the query photograph. Pinterest built a system
which allowed users to hover over pins and find visually similar items
in the catalog~\cite{jing2015visual}. An advantage of these systems is
that they help people find things using an understanding they might
not be able to put into words.

Many lines of research focused on learning the relative similarities
of images. They accomplish this by mapping each image to a
numerical vector, so they can capture the visual similarities in
Euclidean space~\cite{lai2015cvpr, babenko2014eccv, li2016ijcai,
  xia2014aaai, zhu2016aaai}. Using the similar approach but in a
scalable manner, companies have also rolled out their visual search
platforms, from Google Goggles, Google Similar Images, and Amazon Flow
to Microsoft (Bing)~\cite{hu2018kdd}, Pinterest~\cite{jing2015kdd},
eBay~\cite{yang2017kdd}, and Alibaba~\cite{zhang2018kdd}.

All these methods, however, require the user to provide a photograph of
the targeted item. They fail when users do not have an actual visual representation of
the desired item, but instead a mental picture of it. The users
themselves may not know how to properly describe their mental visualization in
words. Our algorithm addresses this issue, as Seeker is able to work with any
embedding representation, including visual, textual and audio.

Moreover, Seeker dynamically adjusts the search results based on
interactive user feedback; all mentioned projects do not allow users
to fine-tune their current-session search with additional
feedback. While we use proprietary embeddings in the examples of this
paper, the underlying engine can operate upon features derived from
other domains (or combinations of domains) as well -- customer
behavior, language understanding, audio, etc.

\section{Problem Formulation}
\label{sec:setting}

\subsection{The Setting}

\begin{figure}[h]
  \includegraphics[scale=0.5]{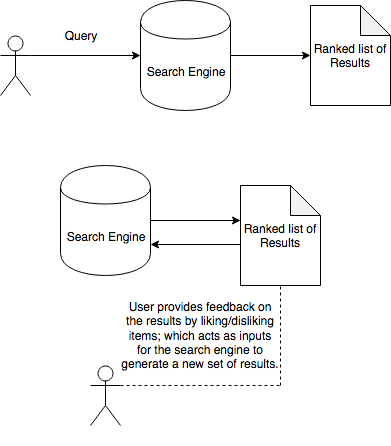}
  \caption{The top figure describes a traditional search engine: A
    user submits a query to the engine and is presented with a
    ranked list of items. Our system (bottom figure) dynamically incorporates
    feedback on results in real-time, generating a new ranked list
    with each like or dislike. }
  \label{fig:setting}
\end{figure}

Figure~\ref{fig:setting} illustrates how Seeker is different from
traditional search. The user starts with a ranked list of results and
provides feedback in the form of likes or dislikes; the search engine
then generates a new set of ranked results, updating the page in real-time.
In this section, we define the notation to formally describe
the above search process.

Assume that we have a catalog of $N$ items, out of which 
$M \ll N$ can be displayed. We model user
feedback as a sequence of likes and dislikes over discretized
timesteps $t_0, t_1, \cdots, t_k$. The user starts with an initial
ranking of items at timestep $t_0$. This initial $t_0$ ranking can
be thought of as Seeker's prior belief on what the user desires,
can be generated from a traditional search or recommendation engine,
and may incorporate diversity or business requirements.

The user interacts with the page by \textit{liking} or
\textit{disliking} items. At each timestep, $t_k$, Seeker produces a
new ranked list of results, based on the feedback from $t_0, \cdots,
t_{k-1}$. It does so by constructing a discrete probability
distribution over the catalog of $N$ items at each timestep. The
probability distribution represents the likelihood of an item being
the user's desired item.

We featurize each catalog item $i$ by embedding it into a vector space
$x_i \in \mathbb{R}^d$. Seeker requires a high correlation between
human perception of similarity and distance metric in the embedded
vector space. Based on the properties of the items displayed,
embedding strategies described in~\cite{le2014distributed, peters2018deep, devlin2018bert, InceptionV3} have been shown to correlate with human perception.

\begin{figure}[h]
  \includegraphics[scale=0.4]{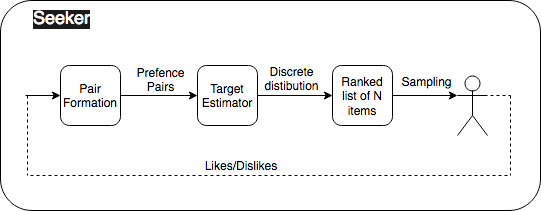}
  \caption{Logical components of Seeker}
  \label{fig:flow}
\end{figure}

Seeker can be divided into three major components, as seen in Figure~\ref{fig:flow}.
Section~\ref{subsec:pairs} describes
 how we convert likes and dislikes to preference pairs and probability
distributions.
 Section~\ref{subsec:target} details how we use preference pairs
to estimate a target's likelihood.
Section~\ref{subsec:BoltzmannExp}
 shows how we use probabilistic sampling to recommend items
 to users at each timestep. 

\subsection{Pairwise Comparison}
\label{subsec:pairs}
Let $a_i^k \in \mathbb{R}^d $ for $i = 1, ..., p$ be the vector
representations of the liked items, and $b_j^k \in \mathbb{R}^d$ for
$j = 1, ..., q $ the vector representations of the disliked items. We
will drop the superscript when the context is clear. Let $A := \{a_1,
..., a_p\}$ and $B := \{b_1, ..., b_q\}$ be the non-empty subsets of
$\{x_1, ..., x_N\}$. We define $s_{ij}$ as the preference pair which
consists of a liked item $a_i$ and a disliked item $b_j$ from sets $A$
and $B$ respectively. We create $pq$ preference pairs from all cross-pairings
between the $p$ likes and $q$ dislikes.

The intuition behind preference pairs follows from our assumption that the
user has some ideal item $t$ (referred to as the target) in their mind
that they wish to find. Then $s_{ij} \in S$ represents the preference
that the user thinks item $i$ is more similar to their desired item
$t$ than item $j$, i.e \textit{they prefer i over j given t}:

\begin{equation} \label{eq:sim}
  || x_i - x_t||^2 < || x_j - x_t||^2.
\end{equation}

Equation~\ref{eq:sim} resolves to item $t$ being spatially closer to
item $i$ than it is to item $j$. In this paper we use the Euclidean
distance to measure vector similarity, but Seeker is agnostic to
the metric used.

We use preference pairs to model the probability of a catalog
item being the hidden target item $t$, featurized as $x_t$.
If we were to present a
user with item $x_i$ and item $x_j$, what is the probability that
they chose $i$ over $j$?  Questions of this form are known as triplets
in the Machine Learning literature~\cite{JainTriplet,schroff2015facenet}.
Equation~\ref{eq:noise} mathematically models our question:
\begin{equation}\label{eq:noise}
	\mathbb{P}\big(s_{ij} | t, i, j\big)= \frac{1}{1 + \exp\big\{-\alpha
      \big( || x_j - x_t||^2 - ||x_i - x_t ||^2\big)\big\}},
\end{equation}
where $\alpha \geq 0$.

Intuitively, the answer to the above triplet question should depend on
how similar items $x_i$ and $x_j$ are to $x_t$. As similarity and
distance are equivalent in our world, the probability of preferring $i$
over $j$ becomes a function of how close $x_i$ and $x_j$ are to $x_t$.
According to this model (and Equation~\ref{eq:noise}),
if items $x_i$ and $x_j$ are
equidistant from the target $x_t$, then they are equally preferred,
and the probability of choosing $i$ over $j$ is
$0.5$. If $x_i$ is the target $x_t$ while $x_j$ is infinitely far away,
the probability of choosing $i$ becomes $1$. For items in the middle
we get a smooth noise model that accounts for the stochasticity in human
decisions.

Our model includes a preference hyperparameter $\alpha$, which
represents our confidence in the vector space representation:
\begin{itemize}
  \item When $\alpha = 0$, then $\mathbb{P}\big( S_{ij} | t \big)=
    0.5$ for all combinations of targets, likes and dislikes. This
    means our embeddings have no correlation with human
    judgment of similarity, and preferring $i$ over $j$ is as good as
    a fair coin flip.
  \item When $\alpha = \infty$, then $\mathbb{P}\big( S_{ij} | t
    \big)= 1$. This removes randomness from the decision process,
    perfectly aligns our representation of human judgment
    with the metric distance, and deterministically picks
    the closer item.
\end{itemize}
We use $\alpha=1$ for the results discussed in Section~\ref{sec:result}.

\section{Item Ranking}
\label{sec:rank}

In this section we describe how we go from preference pairs and a
noise model to a ranked list of items to be displayed to the user.

\subsection{Target Estimation}
\label{subsec:target}

To keep our
notation consistent, we always assume that the user prefers item $x_i$
to $x_j$ when we write $s_{ij} \in S$. We make the further assumption that each preference pair is independent from each other. This is a
simplifying assumption which serves as a good baseline
~\cite{chapelle2011empirical, hill2017kdd}. In Section~\ref{subsec:bipartite}, we investigate ways to drop the independence
assumption. Equation~\ref{eq:likelihood} represents the joint distribution likelihood
of observing preferences $S$, given target item $t$ and likes and dislikes sets $A$ and $B$:
\begin{equation}
  \mathbb{P}(S | t, A, B) = \prod_{s_{ij} \in S} \mathbb{P}\big(s_{ij} | t, i, j \big),
  \label{eq:likelihood}
\end{equation}
where $\mathbb{P}(s_{ij} | t, i, j)$ is defined as in
Equation~\ref{eq:noise}. The log-likelihood becomes:
\begin{equation}
  \log \mathbb{P}(S | t, A, B) = \sum_{s_{ij} \in S} \log \mathbb{P}\big(s_{ij} | t, i, j \big).
  \label{eq:log_likelihood}
\end{equation}

We do not know a priori what is the hidden target $t$. Our goal is to
find $t$ or approximate it. We note that $t$ may not be present in our
catalog, and in this case our goal is to find an item as similar to $t$
as possible. In order to build a probability distribution over our
catalog, we borrow ideas from~\cite{tamuz2011adaptively}. We use the
same noise model, but apply it to recommend items to the user, instead
of learning a metric space. For each catalog item, we compute the
log-likelihood mass of that item being the target, given the user's
likes and dislikes, as shown in Algorithm~\ref{alg:scores}.

\begin{algorithm}
  \caption{Catalog items log-likelihood computation}\label{alg:scores}
  \begin{algorithmic}[l]
    \State scores = [ ]
    \ForAll {items $t= 1, \cdots, N$}
    \State score = 0
    \ForAll {$s_{ij} \in S$}
    \State score += $ \log \mathbb{P}\big(t | s_{ij} \big)$
    \EndFor
    \State scores.append(score)
    \EndFor
  \end{algorithmic}
\end{algorithm}

\subsection{Posterior Construction}
\label{Sec:prior_posterior}
Instead of presenting items according to their likelihood of being the
target, we allow for the inclusion of priors into our model. Let
$\mathbb{P}(i)$ be the prior probability of item $i$ being the actual
desired target. One can compute such priors using traditional search
engines, and personalize them using the user's browsing or purchase
history~\cite{Teo2016airstream}.

Given priors $\mathbb{P}(i)$, the posterior probability of an item
being the target is:
\begin{equation}
  \mathbb{P}(t | S) \propto   \mathbb{P}\big(S | t \big)\mathbb{P}(t),
  \label{eq:posterior}
\end{equation}
and the log-posterior becomes:
\begin{equation}  \label{eq:log_posterior}
  \begin{split}
  \log \mathbb{P}(t | S) &\propto  \log \mathbb{P}\big(S | t \big) + \log \mathbb{P}(t)\\
  & = \sum_{s_{ij} \in S} \log \mathbb{P}\big(s_{ij} | t, i, j \big) +  \log \mathbb{P}(t).
  \end{split}
\end{equation}

At each time step the user provides feedback causing the size of $S$
to grow. Therefore the log likelihood will eventually dominate the
posterior density score. In the early stages when we have fewer likes
and dislikes, our posterior belief on the target is dominated by a
well founded prior. This prevents us from having to wait a long time
before showing meaningful results.

\subsection{Items Recommendation}
\label{subsec:itemRec}

We consider four different ways to display $M \ll N$ items to the user:

\subsubsection{Pure Exploitation/Noiseless}
The simplest approach is to sort the posteriors and recommend the
top $M$ items.
Theoretically, this prevents us from exploring the search space. Practically,
this leads to a poor user experience with limited product diversity.

\subsubsection{Pure Exploration/Random}
The other extreme solution is to show random results all the time,
completely ignoring the posterior densities.

\subsubsection{Epsilon-greedy}
Another approach is to randomize some of the results while leaving the
others untouched, as in Algorithm~\ref{alg:greedy}.  We rank items
by their posterior densities, and replace each item with a random item with
probability $\epsilon$. See~\cite{bubeck2012regret} for a detailed study of $\epsilon$-greedy
algorithms.

\begin{algorithm}
  \caption{Epsilon-greedy sampling}
   \label{alg:greedy}
   \begin{algorithmic}[l]
     \Require $0 \leq \epsilon \leq 1$
     \State  noiseless = argsort \{ $\mathbb{P}(t | S)$ \} in descending order
     \State results = [ ]
     \ForAll {$x_i$, $i \in$ noiseless}
       \State flip $\epsilon$ biased coin
       \If{heads}
         \State $x$ $\sim$ Unif($x_1, \cdots, x_N$)
         \State results = results $\cup$ $x$
       \Else
         \State results = results $\cup$ $x_i$
       \EndIf
     \EndFor
     \State \Return results
  \end{algorithmic}
\end{algorithm}

\subsection{Boltzmann Exploration}
\label{subsec:BoltzmannExp}

The fourth method to recommend items involves sampling without replacement
according to the item's posterior densities. Let $g_j$ be a score
associated with item $j$. A popular way to generate a discrete
distribution over the items is by using the exponential weighing
scheme, known as the softmax or Boltzmann equation:
\begin{equation}\label{eq:softmax}
  p_j = \frac{ e^{g_j}}{\sum_{i = 1}^N  e^{g_i} }.
\end{equation}
Here, $p_j$ is our belief probability that item $j$ is the true
target. Even though $g$ is unconstrained in $\mathbb{R}$, common
values are $g_j = \mathbb{P}(x_j | S)$ and $g_j = \log \mathbb{P}(x_j
| S)$, the latter resulting in polynomial weighing~\cite{RLAlgos,
  Jang2017}.
Note that if the items were equally
spaced, sampling from the discrete distribution $p_j$ is
asymptotically equivalent to sampling from the hidden continuous
distribution, as we show in Appendix~\ref{appendix:sampling}.

Sampling without replacement when $N$ is large can prove to be very
slow. When $N$ and $d$ are large, normalizing our posterior densities
can lead to precision issues with sampling.  We can overcome this
problem by using the \emph{Gumbel-Max} trick~\cite{A*Sampling}, which
shows that adding standard Gumbel noise to $g_i$ and taking the $\max$
is equivalent to sampling according to Boltzmann
(Equation~\ref{eq:softmax}):
\begin{equation}
\argmax_j \{g_j + Gumbel(0,1)\} \sim \frac{ e^{g_j}}{\sum_{i = 1}^N  e^{g_i} }= p_j.
\end{equation}

We sketch the proof for completeness. Let $z_i = g_i +
Gumbel(0,1)$. By the additive property, $z_i \sim Gumbel(g_i, 1)$,
with probability density function (PDF):
\begin{equation}\label{eg:gumbelPDF}
f_i(z) = e^{-\big(z-g_i+e^{-(z-g_i)}\big)},
\end{equation}
and cumulative distribution function (CDF):
\begin{equation}\label{eg:gumbelCDF}
F_i(z) := \mathbb{P} (z_i \leq z) = e^{-e^{-(z - g_i)}}.
\end{equation}

\begin{proof}
Define by $\mathbb{P}(j^*)$ the probability that $z_j$ is the largest among all $z_i$. We have:
\begingroup
\allowdisplaybreaks
\begin{align}
    \mathbb{P}(j^*) & = \int_{z_j=-\infty}^{+\infty} \, f_j(z_j) \, \prod_{i \ne j} \mathbb{P} (z_i \leq z_j) \, dz_j\notag\\
  &= \int_{z_j=-\infty}^{+\infty} \, e^{-\big (z_j-g_j+e^{-(z_j-g_j)}\big )}  \prod_{i \neq j} e^{-e^{-(z_j - g_i)}} \, dz_j\notag\\
    &= \int_{z_j=-\infty}^{+\infty} \, e^{-z_j+g_j-e^{-z_j}\sum_{i = 1}^N e^{g_i}}\, dz_j\\
    &= \frac{e^{g_j - e^{-z_j}\sum_{i = 1}^N e^{g_i}}}{\sum_{i = 1}^N e^{g_i}} \Biggr|_{-\infty}^{+\infty}
    = \frac{e^{g_j }}{\sum_{i = 1}^N e^{g_i}} = p_j. \notag
\end{align}
\endgroup
\end{proof}

Since the added Gumbel noises are independent, showing the $M$ items
with the highest $z_i$ scores is equivalent to sampling $M$ items
without replacement from Equation~\ref{eq:softmax}.

To balance exploration and exploitation, one resorts to
annealing~\cite{aarts1988simulated}, with an appropriately tuned
sequence of learning rate parameters (aka inverse temperature) $\eta_k
\geq 0$ for each timestep $t_k$:
\begin{equation} \label{eq:annealing}
  p_j = \frac{e^{\eta_k\, g_j}}{\sum_{i = 1}^N  e^{\eta_k\, g_i} }.
\end{equation}
Note that $\eta_k=0$ recovers the pure exploration mode, and
$\eta_k=+\infty$ recovers the pure exploitation mode. Varying $\eta_k$
allows us to trade-off exploitation and exploration.

On the other hand, similarly to the proof above, we have
\begin{align}\label{eq:shiftedGumbel}
\argmax_j \{\eta_k\, g_j + Gumbel(0,1)\} &= \argmax_j \{Gumbel(\eta_k\, g_j, 1)\}\notag \\
&\sim \frac{e^{\eta_k\, g_j}}{\sum_{i = 1}^N  e^{\eta_k\, g_i} }.
\end{align}
Note that, by dividing by $\eta_k$, we establish:
\begin{equation}
\argmax_j \{\eta_k\, g_j + Gumbel(0,1)\} \sim \argmax_j \Big \{g_i + \frac{Gumbel(0,1)}{\eta_k}\Big\}.
\end{equation}

Sampling from $Gumbel(\eta_k\, g_j, 1)$ and taking the maximum, as in
Equation~\ref{eq:shiftedGumbel}, is similar to Thompson Sampling in a
bandit setting~\cite{thompson1933likelihood, russo2018tutorial}.  The
crucial difference (and drawback) is that the Gumbel method doesn't 
take into account the uncertainty of the
reward estimates.

Finding the right schedule for $\eta_k$ can be very difficult in
practice~\cite{Vermorel:2005}.  In~\cite{cesa2017boltzmann}, the
authors provide an annealing schedule for $\eta_k$ in a standard
stochastic multi-armed bandit setting, guaranteeing sublinear regret.
Let $n_j$ be the number of times arm $j$ has been played up to
timestep $t_{k-1}$.  For some constant $C>0$, they set $\eta =
\sqrt{n_j/C^2}$, and sample according to:
\begin{equation}\label{eq:boltzmannDoneRight}
\argmax_j \Big \{g_j +  \sqrt{C^2/n_j}\; Gumbel(0,1) \Big\}.
\end{equation}

Equation~\ref{eq:boltzmannDoneRight} decouples the learning rates of the individual items, and
factors-in the uncertainty of the reward estimates. We now have a
proper way to sample from a Boltzmann, with convergence
guarantees.
Even though our setting is not exactly the same
as~\cite{cesa2017boltzmann}, we borrow parts of their sampling
strategy to recommend items to the user. As detailed in the theoretical
justifications of Appendix~\ref{sec:SettingC}, we recommend setting 
$C^2 = 1/8$ for $g_j = \mathbb{P}(x_j | S)$.

As a user can repeatedly interact
with the same item, we treat $n_j$ as the
number of times a user interacts with item $j$. It starts with
$n_j = 1$ and is incremented with every like or dislike to item $j$.
Putting it all together, we obtain our final Boltzmann sampling algorithm
(Algorithm~\ref{alg:boltzmann}).

\begin{algorithm}
  \caption{Boltzmann sampling for recommending items}\label{alg:boltzmann}
  \begin{algorithmic}[l]
    \Require $M \leq N$, $n_i \forall i \in \{1, 2, \cdots, N\}$
     \ForAll{$x_i$, $i= 1, \cdots, N$}
       \State $\gamma_i \sim $ Gumbel$(0, 1)$
       \State $z_i = g_i + \frac{C \, \gamma_i}{\sqrt{n_i}}$
     \EndFor
     \State results = sort(\{ $z_i$ | $i= 1, 2, \cdots, N$ \}) in descending order
     \State \Return top $M$ results
  \end{algorithmic}
\end{algorithm}

\section{Evaluation}
\label{sec:result}

As our experiments require
human judgments, there exist no such ground truth datasets for
validation. Instead we propose an experimental framework with a human
in the loop that simulates the \textbf{Seeker} experience and
generates quantifiable metrics. The evaluation study serves as a
benchmark for future sequential search algorithms.

\subsection{Experimental Setup}
Seeker assumes that the user has a mental image of a target item they
cannot easily express in words. When accessing Seeker,
the user is presented with a subset of $M$ items to interact with,
using like or dislike clicks. At any moment, the user can expand the
catalog listing view by clicking on \lq\lq Explore More\rq\rq. Our
experimental setting mimics this initial user experience.

A single experimental session
involves the following: A user is presented with a target item
$x_t$. This target item is an explicit simulation of the user's hidden
target. At each timestep, we present the user with a grid of $M$
items. The user's goal is to find the target item through a
series of feedbacks. At each timestep, they may like, dislike, or
remove a previously liked/disliked item. Upon receiving user feedback,
we recommend $M$ new items to view in the next timestep.
The session goes on for $K$
timesteps. If the user can find the target within the $M$ items, they
may stop playing. Otherwise, they try to get as close to the
target item as possible based on their perception of similarity.
For our experiments, we set $M=12$, $K=15$, $d=2048$, $N=2228$,
$g_j = \log \mathbb{P}(x_j | S)$, and used an uninformative prior.

We enlisted volunteers to participate in the experiment defined above,
and collected 358 (roughly 90 per sampling algorithm) unique
sessions. The target and exploration algorithm for each session was
selected uniformly at random. Users were instructed to like and
dislike items assuming that they wanted to purchase the target
item. Users had no prior knowledge of the selected catalog or
algorithm. At each timestep, we invoke Seeker to generate a posterior
distribution over the catalog, according to
Equation~\ref{eq:log_posterior}.  This distribution enforces a natural
ranking on the items. We monitor the normalized rank of the target
item at each timestep. The normalized rank $\rho$ is defined as the
rank of the target item divided by the size of the catalog. A target
$x_t$ with a normalized rank of $\rho = 0.1$ means it has a final rank
of $\ceil[\big]{0.1*N}$.

\subsection{Experimental Results}\label{sec:performance}
Seeker aims at helping the user quickly zoom-in on the desired target item.
A typical metric for such recommender systems is
\emph{recall at k}~\cite{evaluatingRecommenderSystems}.
As we have only one target of interest, we measure how close our recommendations
are to target $t$. We can do that using the target's
normalized rank. For a given session $i$, let $\rho_i$ be the lowest
normalized rank attained by $t$ in all $K$ timesteps.
We define \emph{recall @$\rho_j$} as the percentage
of sessions with $\rho_i \leq \rho_j$. For example, a recall of
$0.4$ @$0.02$ means that $40\%$ of sessions achieved a normalized ranking of
$\rho = 0.02$ or less.

\begin{figure}[h!]
  \includegraphics[width=9.5cm]{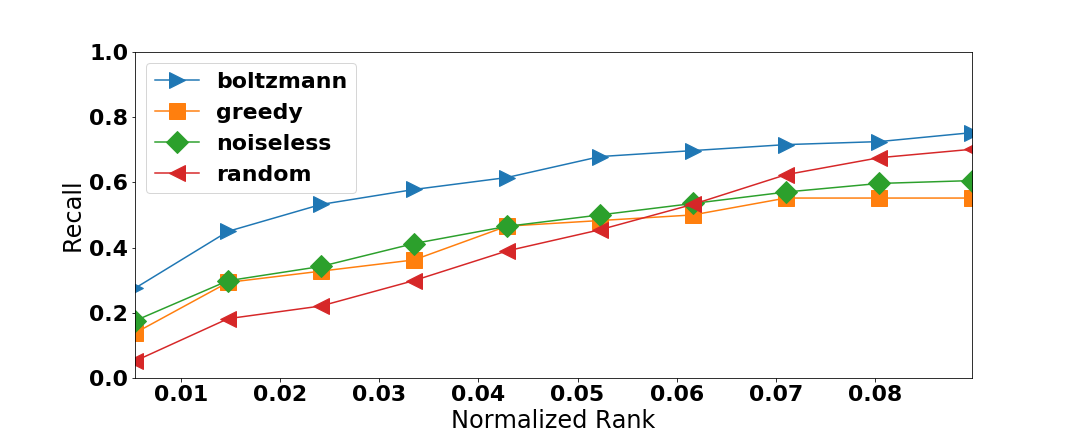}
  \caption{Recall at different normalized rank cutoffs. The plot
    measures how often Seeker ranks the target item better than a given
    percentile within $K=15$
    timesteps.}
  \label{fig:results_recall}
\end{figure}

Figure~\ref{fig:results_recall} plots recall @$\rho$
for our sampling strategies. We plot $\rho$ up to $0.1$,
as the user is unlikely to scroll past higher percentiles.
Boltzmann exploration achieves the highest recall, dominating all other strategies.
Noiseless and Greedy perform similarly, outperforming random at lower recalls.
Random improves at higher recalls due to its higher degree of exploration,
where the target gets ranked high by pure chance.

\begin{figure}[h!]
  \includegraphics[width=9.5cm]{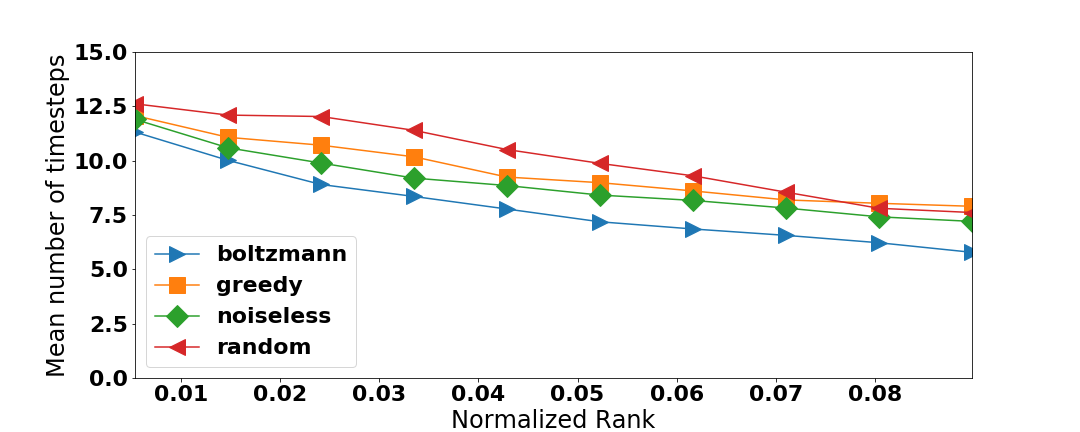}
  \caption{Mean number of timesteps (interactions) until target item is ranked better
    than a given percentile.}
  \label{fig:results_timesteps}
\end{figure}

Figure~\ref{fig:results_timesteps} plots the convergence time
of our sampling strategies. From the user's perspective,
this reflects how long it takes to find a reasonably close
approximation of the target item.
We report the mean number of steps it takes for
the rank of the target item to drop below a given recall cutoff $\rho$.
Boltzmann exploration consistently outperforms the other strategies.
Greedy and Noiseless surpass Random, but their advantage diminishes at
higher rank cutoffs.

\subsection{Discussion}\label{sec:discussion}
We would like to point out that the experimental setup described above
is not restrictive. Although we do present a window of $M$ items with which
the user interacts, the user can expand the window size $M$
by explicitly clicking on an \lq\lq Explore More\rq\rq\ option. Once in
the expanded view, scrolling down past the last displayed item
triggers the display of additional items in an infinite scroll mode
covering the whole $N$ catalog items. Since we maintain an explicit
ranking on all items, this mode of experimentation merges naturally
with our algorithm.

Infinite scroll, such as home feeds on social media
websites like Facebook and twitter, may create a better user
experience and allow for the user to browse quickly. But when the target
is explicit, such an infinite scroll feature makes our experimental
framework trivial -- the user can just scroll until they find the
target. This prevents us from gaining insight about convergence, which
explains why we limited our study to the windowed-version of the
application. We consider our experimental setup a restrictive 
experience in terms of user experience.

Additionally, the catalog contains multiple similar items.
This leads to a large number
of identical feature-vector representations, making it challenging to 
surface the
target item among $M=12$ items in just $K=15$ timesteps.
Hence, it is likely that the Section~\ref{sec:performance}
experimental results are pessimistic, 
as users are constrained from browsing the search space
efficiently. Nevertheless, as the top-most items get
the most visibility, Seeker's ability to quickly zoom-in to 
the item of interest
remains crucial.

On occasions, the Seeker interface produces pages with very similar
items ranked closely, leading to lack of exploration. Two items which
look mostly identical are likely to have similar vector 
representation and hence may appear adjacent to
each other~\cite{nassif2016music}. In a deterministic setting, this
would have resulted in a page full of very similar items, and
prevented the user from pivoting to other parts of the
catalog. Although Boltzmann exploration offers a principled remedy,
depending on the use case, one may want to model additional uncertainty
into the user actions in the early stages of the interactions.  As a
remedy, we can modify the posteriors by adding noise, using submodular
functions~\cite{InteractiveSubmodularBandit}, or determinental point
processes~\cite{affandi2012markovDPP}.

The constant $C$ in Algorithm~\ref{alg:boltzmann} is
borrowed from the work in~\cite{cesa2017boltzmann} which uses the
non-contextual stochastic multi-armed bandit setting. Under their setting
$C$ is a reasonable estimate to bound variance. However, items in our
search space have features that are shared and correlated. Our
sampling strategy currently does not take into account this covariance
properly when making recommendations. We leave augmenting our sampling
algorithm with a new variance bound for future work.

\section{Future Work}
\label{sec:future}

We are considering improving this work on multiple fronts.

\subsection{Bipartite Preference Model}
\label{subsec:bipartite}

When constructing the preferences from user feedback in
Section~\ref{subsec:pairs}, we treat the likes and dislikes independently,
valuing them equally. However, intuitively, if we
put more emphasis on likes, we may be able to find the target
faster. Likes are less ambiguous than dislikes:
likes have a clearer implication when treated in
isolation, while dislikes usually require context to be a useful
learning signal. In fact, we empirically
observe that the likes are more clustered with one another than
the dislikes to themselves.

\begin{figure}[h]
\centering

\definecolor{myblue}{RGB}{80,80,160}
\definecolor{mygreen}{RGB}{80,160,80}

\begin{tikzpicture}[thick,
  every node/.style={draw,circle}, fsnode/.style={fill=myblue},
  ssnode/.style={fill=mygreen}, every fit/.style={ellipse,draw,inner
    sep=-2pt,text width=2cm}, ->,shorten >= 3pt,shorten <= 3pt ]

\begin{scope}[start chain=going below,node distance=7mm]
\foreach \i in {1,...,5}
  \node[fsnode,on chain] (f\i) [label=left: $a_\i$] {};
\end{scope}

\begin{scope}[xshift=4cm,yshift=-0.5cm,start chain=going below,node distance=7mm]
\foreach \i in {1,...,4}
  \node[ssnode,on chain] (s\i) [label=right: $b_\i$] {};
\end{scope}

\node [myblue,fit=(f1) (f5),label=above:$A$] {};
\node [mygreen,fit=(s1) (s4),label=above:$B$] {};

\foreach \i in {1,...,5}
	\foreach \j in {1,...,4}
		\draw (f\i) -- (s\j);
\end{tikzpicture}
\caption{A complete directed bipartite graph between the sets of likes and dislikes.}
\label{fig:bipartite_graph}
\end{figure}
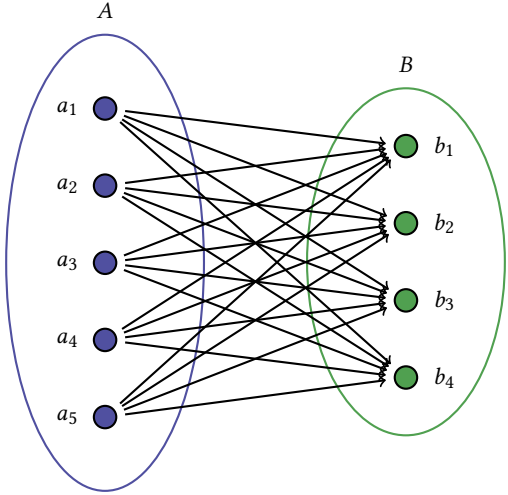

We can mathematically model the emphasis of likes over dislikes by
assuming the likes are independent of each other but the dislikes are
conditioned on the likes. Conditional dependencies can be expressed in
the form of a bipartite graph as shown in Figure~\ref{fig:bipartite_graph}.

Using Bayes rule, we can represent Figure~\ref{fig:bipartite_graph} as:
\begin{equation}
	\mathbb{P} (S | t) = \prod_{i = 1}^p \mathbb{P} (a_i | t) \prod_{j = 1}^q \mathbb{P} (b_j | a_1, \cdots, a_p, t).
\end{equation}
Here we assume that:
\begin{equation}\label{eq:like}
	\mathbb{P} (a_i | t) = \frac{1}{\exp \big\{ \alpha_1 ||a_i - x_t ||^2 \big\}},
\end{equation}
and
\begin{multline}\label{eq:dislike}
  \mathbb{P} (b_j | a_1, \cdots, a_p, t) = \\
  \frac{1}{1 + \exp \Big\{ - \alpha_2 \Big( ||b_j - x_t ||^2 -  \min\limits_{i \in \{1, \cdots, p\}} ||a_i - x_t||^2 \Big) \Big\}},
\end{multline}
with $\alpha_1, \alpha_2 \geq 0$.

We interpret the model in the following way. Equation~\ref{eq:like}
conveys that the probability of liking
item $i$ is proportional to how similar $i$ is to target $t$.
Equation~\ref{eq:dislike} conveys that
the probability of disliking item $j$ is proportional to its relative distance
to the target as compared to the relative distance between the target and liked items $A$. One can quantify the distance between $t$ and $A$ in different
ways. Here we propose $min$ to reflect the customer's gradual approach towards the target.
We leave the evaluation of this model to future work.

\subsection{Incorporating Additional Feedback}
\label{subsec:otherfeedback}

So far, the only form of feedback that the user provides is in the
form of likes and dislikes. Consider the situation where the user
provides feedback in the form of a text or utterance. This
transitions us into the guided conversational search paradigm and we
could incorporate some of the strategies described 
in~\cite{huang2018flowqa, wen2016network, shah2018building}.

Assume we have a technique (like LSTM to create word embeddings) 
to map a spoken feedback into a
vector $r_i \in R^d$. We want to incorporate this feedback into the
model. Equation~\ref{eq:posterior} becomes:
\begin{equation}
\mathbb{P} (t | S, \mathbf{r}) \propto \mathbb{P} (S | t, \mathbf{r}) \mathbb{P} (t | \mathbf{r}),
\end{equation}
where
\begin{equation}
	\mathbb{P} (t | \mathbf{r}) = \frac{1}{1 + \exp \Big\{ - \beta \sum\limits_k r_k^T x_t \Big\}}.
\end{equation}
To estimate $\mathbb{P} (S | t, \mathbf{r})$, we change Equation~\ref{eq:noise} to:
\begin{multline}
\mathbb{P}\big( s_{ij} |  t, i, j, \mathbf{r} \big)= \\
\frac{1}{1 + \exp \Big\{-\alpha  \Big( || x_j - x_t||^2 - ||x_i - x_t ||^2 + \sum\limits_k r_k^T (x_i - x_j) \Big)\Big\}}.
\end{multline}

Although we used text/speech as an example, the additional feedback embedding $r$ can originate from an arbitrary source. Similarly, we can incorporate extra feedback into the bipartite preference model of Section~\ref{subsec:bipartite}.

\subsection{Personalized Recommendations}
Another possible direction is to personalize Seeker. 
Let $c$ be an
embedding vector for each user. The dataset now comes in
the form of quadruplets $(c, t, a, b)$, where each user $c$ has a
target $t$ and pairs $(a, b)$ of likes and dislikes.

To personalize, we define a synthetic embedding
kernel $\phi(c, x)$, where $x$ denotes an item. For example, we can use
element-wise product:
\begin{equation}
	\phi(c, x) = c \odot x.
\end{equation}
Now, we can substitute this kernel into our modeling formulas,
replacing any item $x$ with personalized embedding $\phi(c, x)$.

\section{Conclusion}
\label{sec:conclusion}

This paper presents Seeker, an interactive, real-time search system.
Seeker allows users to search for products even when it is difficult
to describe them in words. Unlike embedding-based search engines,
this method does not require a preknown representation of the desired
item. With interactive binary feedback, our system learns to
dynamically refine search results from the user's preferences in real
time. Our evaluation results show that
our Boltzmann exploration method allows users to find their products
more quickly and with greater regularity compared to alternative
exploration strategies. 

\begin{acks}
We gratefully acknowledge Kevin Jamieson and Lalit Jain for sharing
their experience of designing adaptive algorithms. We would like to
thank Miguel Jimenez Gomez, Xiaopeng Zhang and Andrea Matsunaga for
their software development expertise. In addition, we thank the
volunteers for their help in performing the user study. Finally, we
thank Amazon for the opportunity to conduct this research project.
\end{acks}

\bibliographystyle{ACM-Reference-Format}
\balance
\bibliography{bayesian_muse}

\appendix
\section{Asymptotic Sampling Equivalence}
\label{appendix:sampling}

Given $T$ as a compact (i.e. closed and bounded) subset of $\mathbb{R}^d$. Let $f: T \to \mathbb{R}_+$ be a continuous probability density function. Let the set $X_n = \{x_1, ..., x_n\}$ consist of points $x_i$'s that are equally spaced on $T$ in the grid-like manner such that $\sum_{i = 1}^n f(x_i) > 0$. Consider the following two ways of sampling over $X_n$:

\begin{enumerate}
	\item Each time, sample $x$ from $f$ on $T$, and choose $x_k \in X_n$ if and only if $x_k = \argmin_{x_i \in X_n} d(x, x_i)$, where the metric $d(\cdot, \cdot)$ is usually the Euclidean metric. We assume argmin is unique. 
	
	\item Each time, sample each $x_k \in X_n$ from the discrete distribution on $X_n$ so that $x_k$ is chosen with probability \mbox{$f(x_k) / \sum\limits_{i = 1}^n f(x_i)$}.
\end{enumerate}

Define:
\begin{equation}
	D_n := \sum_{x_k \in X_n} \big| \mathbb{P} (x_k | \text{ Method 1}) - \mathbb{P} (x_k | \text{ Method 2})\big|.
\end{equation}
Prove that $\displaystyle \lim_{n \to \infty} D_n \to 0$.

\begin{proof}

Partition $T$ into $n$ disjoint regions $(J_1, ..., J_n)$ in the grid-like manner such that for each $i \in \{1, ..., n\}$ for each $x \in J_i$,
\begin{equation*}
	i = \argmin_{k \in \{1, ..., n\}} d(x, x_k).
\end{equation*}
Since the points $x_i$'s are equally spaced on $T$, the regions $J_i$'s all have the same measure: $m(J_i) = m = \frac{m_T}{n}$, where $m_T$ is the (fixed) Lebesgue measure of $T$. So
\begin{equation*}
	\mathbb{P} (x_k | \text{ Method 1}) = \mathbb{P} (x \in J_k) = \int_{J_k} f(x) dx = m f(x_k^*) \text{ for all } k.
\end{equation*}
Here the first equation holds by the definition of $J_i$'s, and the second by the Mean Value Theorem (MVT) for some $x_k^* \in J_k$. On the other hand,
\begin{equation*}
	1 = \int_T f(x) dx = \sum_{i = 1}^n \int_{J_i} f(x) dx = \sum_{i = 1}^n m f(x_i^+),
\end{equation*}
where the second equation holds by the additivity of integral, and the last equation holds by the MVT for some $x_i^+ \in J_i$.

Since $f$ is continuous on the compact subset $T$ of $\mathbb{R}^d$, there is an upper bound $U$ such that $f(x) < U \text{ for all } x \in T$. Moreover by the Heine - Cantor theorem, $f$ is uniformly continuous on $T$. 

Now fix $\epsilon > 0$, $\epsilon < 1/2$. By uniform continuity of $f$ on $T$, there exists $\delta > 0$ such that for all $x_1, x_2 \in T$ with $d(x_1, x_2) < \delta$, we have $|f(x_1) - f(x_2)| < S\epsilon$ where
\begin{equation*}
	S = \min \left(\frac{1}{2m_T (1 + U m_T + m_T)}, 1, \frac{1}{m_T}\right).
\end{equation*}

Because the regions $J_i$'s are partitioned in the grid-like manner, there exists $N_0 \in \mathbb{Z}_+$ such that for all $n > N_0$, the diameter of each $J_i$ is smaller than $\delta$.
This implies $d(x_i^+, x_i) < \delta$ for all $i$ and $d(x_k^*, x_k) < \delta$ for all $k$. Hence for all $n > N_0$, we have
\begin{equation*}
	|f(x_i^+) - f(x_i)| < S\epsilon \text{ and } |f(x_k^*) - f(x_k)| < S\epsilon \text{ for all } i, k \in \{1, ..., n\},
\end{equation*}
which implies:
\begin{equation*}
	\left|m \sum_{i = 1}^n f(x_i) - 1\right| = \left|m \sum_{i = 1}^n f(x_i) - m \sum_{i = 1}^n f(x_i^+)\right| < mn S\epsilon = m_T S\epsilon.
\end{equation*}
Therefore for each $k \in \{1, 2, ..., n\}$,  
\begin{equation*}
\begin{split}
	& \left|\mathbb{P} (x_k | \text{ Method 1}) - \mathbb{P} (x_k | \text{ Method 2})\right| = m\Bigg| f(x_k^*) - \frac{f(x_k)}{m \sum\limits_{i = 1}^n f(x_i)}\Bigg| \\
 	& = \frac{m_T}{n}\Bigg| f(x_k) + t_1 - \frac{f(x_k)}{1 + t_2} \Bigg|, \text{ where } |t_1| < S\epsilon, |t_2| < m_T S \epsilon \\
 	& = \frac{m_T}{n|1 + t_2|} \left| (f(x_k) + t_1)(1 + t_2) - f(x_k) \right|\\
  & = \frac{m_T}{n|1 + t_2|} |t_1 + f(x_k) t_2 + t_1 t_2| \\
  & \leq \frac{m_T}{n|1 + t_2|} \left(|t_1| + |f(x_k)| |t_2| + |t_1| |t_2|\right)\\
& < \frac{m_T}{n|1 + t_2|} (1 + U m_T + m_T S \epsilon) S\epsilon \\
	& < \frac{2m_T}{n} (1 + U m_T + m_T) S\epsilon \; \; \;  (\text{because } S\epsilon < 1, m_T S\epsilon < 1/2).
\end{split}
\end{equation*}
This implies that for all $n > N_0$, 
\begin{equation*}
	D_n < 2m_T (1 + U m_T + m_T) S\epsilon \leq \epsilon.
\end{equation*}
This ends the proof.
\end{proof}

\section{Setting Parameter $C$}
\label{sec:SettingC}

Given an item, a user can
like or dislike it. Our rewards are thus binary, making the reward
distribution $1/2$-subgaussian with variance factor $\sigma^2 = 1/4$.
We follow~\cite{cesa2017boltzmann}'s Theorem 3 computations with 
a standard Gumbel noise  $Gumbel(0,1)$ (see Equations~\ref{eg:gumbelPDF} 
and~\ref{eg:gumbelCDF}). We do not introduce extra variable $c$ in 
the proof of Lemma 3, setting $L = \frac{9C^2\log^2_+ (T \Delta_i^2)}{\Delta_i^2}$.
We thus bind the regret $R_T$ as:
\begin{equation}
  R_T \le
  \sum_{i=2}^N \frac{9C^2 \log^2_+(T\Delta_i^2)}{\Delta_i}
  +
  \sum_{i=2}^N \frac{36 C^2e^{\sigma^2/2C^2} }{\Delta_i} + \sum_{i=2}^N \Delta_i.
\end{equation}
Here the finite horizon $T$ is the final timestep, and the gap
$\Delta_i$ is the difference between the mean reward of the optimal
item, and the mean reward of item $i$.

Although $T$ may potentially be specified, $\Delta_i$ is unknown. To obtain a
small regret, the authors recommend setting $C = \sigma$. But one can
easily see that choosing $C = \sigma/\sqrt{2}$ leads to an even
smaller regret. We therefor set $C^2 = \sigma^2/2 = 1/8$.

\end{document}